# Interpreting the HI 21-cm cosmology maps through Largest Cluster Statistics - I: Impact of the synthetic SKA1-Low observations

**Saswata Dasgupta,**[a] **Samit Kumar Pal,**[a] **Satadru Bag,**[b,c] **Sohini Dutta,**[a] **Suman Majumdar,**[a,d] **Abhirup Datta,**[a] **Aadarsh Pathak,**[e] **Mohd Kamran,**[f] **Rajesh Mondal,**[g] **Prakash Sarkar**[h]

[a]Department of Astronomy, Astrophysics & Space Engineering, Indian Institute of Technology Indore, Indore 453552, India
[b]Korea Astronomy and Space Science Institute, Daejeon, Republic of Korea
[c]Department of Physics, Technical University Munich, James-Franck-Str. 1, 85748 Garching, Germany
[d]Department of Physics, Blackett Laboratory, Imperial College, London SW7 2AZ, U. K.
[e]School of Physics, University of Melbourne, Parkville, VIC 3010, Australia
[f]Theoretical astrophysics, Department of Physics and Astronomy, Uppsala University, Box 516, 751 20 Uppsala, Sweden
[g]School of Physics and Astronomy, Tel-Aviv University, Tel-Aviv, 69978, Israel
[h]Department of Physics, Kashi Sahu College, Seraikella, Jharkhand - 833219, India

E-mail: saswata.iiti@gmail.com

**Abstract.** We analyse the evolution of the largest ionized region using the topological and morphological evolution of the redshifted 21-cm signal coming from the neutral hydrogen distribution during the different stages of reionization. For this analysis, we use the "Largest Cluster Statistics" - LCS. We mainly study the impact of the array synthesized beam on the LCS analysis of the 21-cm signal considering the upcoming low-frequency Square Kilometer Array (SKA1-Low) observations using a realistic simulation for such observation based on the 21cmE2E-pipeline using OSKAR. We find that bias in LCS estimation is introduced in synthetic observations due to the array beam. This in turn shifts the apparent percolation transition point towards the later stages of reionization. The biased estimates of LCS, occurring due to the effect of the lower resolution (lack of longer baselines) and the telescope synthesized beam will lead to a biased interpretation of the reionization history. This is important to note while interpreting any future 21-cm signal images from upcoming or future telescopes like the SKA, HERA, etc. We conclude that one may need denser $uv$-coverage at longer baselines for a better deconvolution of the array synthesized beam from the 21-cm images and a relatively unbiased estimate of LCS from such images.

# Contents

# 1 Introduction

The two least understood eras in our universe's history are the Cosmic Dawn (CD) and the subsequent Epoch of Reionization (EoR). The Cosmic Dawn (CD) began when the Cosmic Dark Ages came to an end and the first luminous sources began forming. The Intergalactic Medium (IGM), which was primarily made up of neutral hydrogen ( H I ), began to get ionized as a result of the high-energy photons that emanated from these sources. This time period is known as the Epoch of Reionization (EoR). Our comprehension of the "phase transition" of baryon distribution in the universe from the epoch of the last scattering to the present era is hampered by the absence of observations of this epoch [see e.g. 1–3]. In theory, taking direct images of these epochs with various telescopes is the ideal method for studying them. One of the main scientific objectives of the fairly young James Webb Space Telescope (JWST) is to observe the first sources of light during the EoR. However, until a statistically significant number of such observations are done, the inference from these observations will remain skewed in favour of the brightest sources [4]. In order to explore this epoch, many complementary probes, particularly those involving hydrogen, the most common baryonic element present in the Universe, are used. By examining the Thompson scattering of CMBR photons off free electrons [5, 6], the luminosity function and clustering properties of the Ly$\alpha$ emitters [7–9] and the Lyman-emitting high redshift quasars [10–13], it has been determined that the end of reionization occurred at a redshift of $z \sim 6$. However, these observations miss out on a crucial part, that is the three-dimensional distribution of H<small>I</small> in the IGM at different stages of reionization. Thus, including topological analysis of the distribution of the regions of ionized hydrogen ( H<small>II</small> ) in the IGM will provide crucial inputs that can further add to the inference on the ionizing sources (probed via e.g. JWST-like telescopes) and their impact on the IGM.

The 21-cm line of neutral hydrogen caused by the spin-flip transition of an electron in the ground state from a parallel to an anti-parallel state has the potential to answer a plethora of unresolved questions related to the EoR as it directly probes neutral hydrogen [1, 3]. It is expected to be especially instrumental in tracking the evolving state of the IGM and via that the reionization history. One of the major obstacles to the observation of this signal is the foreground it is buried in. Since the galactic and extra-galactic foregrounds are $\sim 10^5$ times brighter [14–20] than the feeble 21-cm signal, they easily obscure it and create hindrances in its detection [21]. The capacity to detect the signal is also further diminished by a variety of additional factors, including instrument noise [22, 23], telescope beam effect, and effect of the ionosphere [e.g.[24, 25], Pal et al., in prep ]. The upgraded Giant Metrewave Radio Telescope (uGMRT)[1] [26], the Low Frequency Array (LOFAR) [2] [27], the Murchison Widefield Array (MWA)[3] [28], the Precision Array for Probing the Epoch of Reionization (PAPER) [4] [29], and the Hydrogen Epoch of Reionization Array (HERA) [5] [30, 31] are examples of modern radio telescopes that are presently aiming to detect the 21-cm signal via the statistical fluctuations in the signal in Fourier domain and constrain its spherically averaged power spectrum. The Square Kilometre Array's [6] [32, 33] low-frequency component (SKA1-Low), however, will also be sensitive enough to produce images of the 21-cm signal. The EoR can be analysed as a series of images, each from a different redshift, by observing over a range of frequencies. 21-cm tomography is the study of these images at various frequencies (or redshifts) [34]. We will be able to study the evolution of the signal and, consequently, the development of reionization using these tomographic data sets.

It is only reasonable to employ Fourier statistics, such as the power spectrum [27, 35–47] and the multi-frequency angular power spectrum [48–51], as the target statistic for the detection, since the basic observable in these interferometers is visibility, which is nothing but the Fourier transform of sky brightness convolved with the primary beam [52]. Due to their low sensitivity, the first-generation telescopes have been able to produce only upper limits on the signal power spectra. However, the power spectrum only offers a thorough statistical interpretation for a Gaussian random field. The interaction between the underlying matter density and the changing distribution and sizes of the ionized regions determines the fluctuations in the EoR 21-cm signal. This complex origin makes the signal highly non-Gaussian [43, 53–56]. This non-Gaussianity, cannot be quantified through power spectrum and thus higher-order statistics are needed for its proper quantification. Bispectrum [55, 57–64] and Trispectrum [65] are two higher order statistics which have been explored in this context. However, one should note that even these higher-order Fourier statistics do not contain the phase information of the 21-cm signal fluctuations.

---

[1] http://www.gmrt.ncra.tifr.res.in
[2] http://www.lofar.org/
[3] http://www.mwatelescope.org/
[4] http://eor.berkeley.edu/
[5] https://reionization.org/
[6] http://www.skatelescope.org/

As the future SKA1-Low is expected to produce high-resolution tomographic images of the EoR, a significant amount of work has already been done to find the optimal image analysis method to extract a significant amount of information from such future observations. These methods have been mostly tested using simulated tomographic images. These methods mainly focus on tracking and analyzing the evolving topology and morphology of the 21-cm field during the EoR. A good quantum of work related to analysis of 21-cm images has been done via Minkowski Functionals (MFs) [66–71] and Minkowski Tensors [72]. There are various other methods which are based on percolation theory that tracks the abrupt change in the topology of the 21-cm field [70, 71, 73–75]. Along with these, granulometry [76] and persistence theory [77] has also been used for analyzing the topological phases of ionized hydrogen ( $H_{II}$ ) regions during the EoR. Analyses involving the theory of Betti numbers [78, 79] and local variance [80] focus on the topology like the connectivity, genus and surface area to quantify the IGM state.

However, it is commonly accepted that the conclusions obtained using these approaches rely on the detection of numerous $H_{II}$ regions at any stage with a wide range in their sizes. The study done by Bag et al. [70] shows that the detection of only the largest ionized region is sufficient to draw inferences on the percolation process. They use a novel statistic named the "Largest Cluster Statistics (LCS)" along with Shapefinders to draw this conclusion. Pathak et al. [81] have taken this tool a step further and have distinguished the different major reionization scenarios (inside-out and outside-in) through the same analysis.

In this work, we test the ability of the LCS to extract optimal information from the fluctuating 21-cm field under a realistic observation scenario. For this, we first reduce the resolution of our simulated 21-cm maps to mimic the telescope resolution. Next, they are corrupted with Gaussian random noise and then smoothed out with a Gaussian kernel as per the specifications of a SKA1-Low observation. To test the robustness of this analysis further we simulate a mock SKA1-Low observation, starting from the distribution of its station till the final visibility output, using the 21cmE2E-pipeline [82] which utilizes the OSKAR software. Afterward, we create the final images using the CASA CLEAN algorithm from these visibility data sets. Finally, after obtaining the clean images, we use the *SURFGEN2* code [71, 83] to calculate the LCS and draw inferences on the percolation process which tells us about the sudden merger of the ionizing regions during the EoR.

This paper is structured as follows: In Section 2, we discuss the simulated EoR 21-cm maps. Additionally, this section briefly discusses the 21cmE2E pipeline and the different CLEAN algorithms employed using CASA to obtain simulated observational maps. Section 3 comprises the method of binarization of the 21-cm maps. In Section 4, we discuss the results obtained and what inferences can be drawn regarding the EoR. We summarize our results and conclude in Section 5.

The cosmological parameters from the *Wilkinson Microwave Anisotropy Probe* (WMAP) five-year data release have been used throughout the paper which details as follows: $h = 0.7$, $\Omega_m = 0.27$, $\Omega_\Lambda = 0.73$, $\Omega_b h^2 = 0.0226$ [84].

## 2 Simulating the SKA observations of the EoR

### 2.1 Semi-numerical simulation of reionization

Simulation of the reionization era from the first principle is quite an expensive task in terms of resources. To account for the large-scale density fluctuations of matter, the simulation has to be done in a large cosmological volume ($\sim 1 \, \text{Gpc}^3$). In order to correctly mimic the properties of the reionizing sources, which are typically galaxies smaller than $\sim 10 \, \text{kpc}$, it is also required to identify these sources and mimic their internal physical processes. This requires a high dynamic range in terms of mass and length scale in these simulations.

The complete intricacy of radiative transfer across the clumpy IGM can only be accounted for by the numerical solution of the cosmic radiative transfer equation [74, 85–88] along the path of every ionizing and heating photon. On the other hand, radiative transfer simulations are still computationally expensive, thus the equation is typically solved using suitable approximations by calculating the ionized hydrogen's temporal evolution [54, 89]. One such method is employed by the radiative transfer algorithm "Conservative Causal Ray-tracing methodology" ($C^2$-RAY), which functions by following rays from all sources and iteratively solves for the time evolution of the ionized hydrogen fraction ($\bar{x}_{\text{HII}}$).

It is computationally very expensive to use these simulations to explore the vast multi-dimensional reionization parameter space. Thus semi-numerical simulations [2, 39, 40, 43, 90–92] can be used to account for these limits as a realistic trade-off. Based on the excursion set formalism [93], the semi-numerical technique is computationally effective. One of these approaches starts by creating the dark matter density field using perturbation theory, then calculating the collapsed fraction in each grid cell (of size R and density fluctuation $\delta$) using the analytical expression for the conditional mass function, followed by creating the ionization field using the excursion set formalism [91]. An alternative tactic is to run a full dark matter-only $N$-body simulation and apply a suitable group-finder technique to discover the halos [2, 90]. In this work, we have used the 21-cm maps that were generated using semi-numerical simulations presented in [40, 41].

The simulation combines three main steps: an N-body gravity-only simulation to generate the dark matter field, a Friends-of-Friends (FoF) halo finder to identify collapsed objects in this field, and an excursion set formalism to generate the ionization field using the dark matter and the halo fields.

Initially, the N-body gravity-only simulation is used to model the distribution of dark matter in the Universe at a set of given redshifts. The Friends-of-Friends (FoF) halo finder algorithm is employed on this dark matter field to identify and track the evolution of gravitationally bound dark matter halos in this field.

Finally, The ionization field is generated using the a modified excursion set formalism. In our simulation it is implemented in the following manner: we consider that the sources of ionizing photons are hosted by the dark matter halos. The further assumption is that the number of ionizing photons produced by a halo is proportional to it's mass. One can in principle adopt an even more complex relationship between halo mass and ionizing photon production rate which will include the scatter in photon production rate due to the effect of feedback in star forming galaxies and various other physical processes that goes on inside the first sources of light. For the sake simplicity, here we consider a simple proportionality relationship between the halo mass and the ionizing photon production rate. The proportionality constant $N_{ion}$ effectively represent a combined effect of various degenerate physical parameters of this problem e.g. star formation rate, fraction of baryons that are converted into stars, escape fraction of ionizing photons etc. The other parameter that determines whether a halo will be hosting any photo ionizing sources or not is the minimum halo mass $M_{hmin}$. We assume that all halos above this minimum mass hosts ionizing photon sources and produces ionizing photons that is proportional to their mass.

In our semi-numerical scheme of ionization we next ask the question whether a region in the IGM has enough photons to self-ionize or not. To answer this point we need to know how far away a photon can travel from it's point of origin. This third criteria is manifested through the parameter $R_{mfp}$, which is mean free path of ionizing photons. Next we assume that the hydrogen follows the underlying dark matter distribution, this helps us to create a mock hydrogen distribution from the dark matter field with a bias factor 1. We create a mock ionizing photon field by multiplying the $N_{ion}$ with the already existing halo field. Next, following the excursion set approach we smooth the neutral hydrogen field and the ionizing photon field, starting with a sphere of radius as small as the resolution of the simulation (0.56 Mpc) to as large as the $R_{mfp}$. At each stage of smoothing we go to each pixel of the simulation and ask the question if the number of ionizing photons at that pixel is greater than the number of hydrogen atoms in that pixel. For any of the smoothing radius within the above mentioned range if this ionization criteria is satisfied, we then flag that pixel to be ionized. The pixels in which this criteria is never satisfied for any of the smoothing radius in the above mentioned range, we then flag those pixels to be partially ionized, where the partial ionization fraction in that pixel is determined by the ratio, the number of ionizing photons in the pixel for the smallest smoothing radius divided by the number of neutral hydrogen in that pixel for the same smoothing radius. The ionization field produced in this fashion is then converted into the 21-cm brightness temperature field following equation (2.1) discussed in detail in the next section. The specific values of these three key parameters that we have used for our simulation presented here are mentioned in Section 2.2.

## 2.2 Simulated 21-cm fields from the EoR

The H<small>I</small> 21-cm signal is observed in contrast with the Cosmic Microwave Background Radiation (CMBR) quantified using the differential brightness temperature ( $\delta T_b$ ) and is expressed as:

$$\delta T_b \approx 27 x_{HI}(1+\delta) \left(\frac{1+z}{10}\right)^{\frac{1}{2}} \left(1 - \frac{T_{CMB}(z)}{T_S}\right) \left(\frac{\Omega_b}{0.044} \frac{h}{0.7}\right) \left(\frac{\Omega_m}{0.27}\right)^{-\frac{1}{2}} mK \qquad (2.1)$$

The neutral fraction of hydrogen is denoted by $x_{HI}$, while the density fluctuation is denoted by $\delta$. The CMB temperature at a redshift of $z$ and the spin temperatures of the two states of hydrogen are, respectively, denoted by $T_{CMB}(z)$ and $T_S$. It can be clearly seen from equation 2.1 that, when $T_{CMB} \approx T_S$, no signal can be detected. The 21-cm signal can be detected against the background of the CMB because, according to the Wouthuysen Field effect, the spin temperature will get coupled to the gas temperature during EoR. When the gas temperature is higher than $T_{CMB}$, the signal can be detected in emission and when the gas temperature is lower than $T_{CMB}$, it is seen in absorption. In our realization, we assume a high spin-temperature limit ($T_S \gg T_{CMB}$). Since even relatively small amounts of X-ray radiation can cause the gas temperature to rise above the CMB temperature, it is generally accepted that this is a reasonable assumption for the period when reionization is well underway. The frequency axis is one of the three axes in this three-dimensional data set, while the other two are assumed to indicate the position in the sky. However, in this work, we use simulated coeval signal cubes rather than light cone cubes. It is also important to note that, initially in our simulated signal cubes pixels having a value $\delta T_b = 0$ are ionized pixels. The ionization state for a particular redshift is quantified using the mass-averaged neutral fraction ($\bar{x}_{\rm HI}$) and the Filling factor (or volume-averaged ionization fraction). Here, the mass averaged neutral fraction ($\bar{x}_{\rm HI}$) is defined as:

$$\bar{x}_{\rm HI}(z) = \bar{\rho}_{\rm HI}(z)/\bar{\rho}_{\rm H}(z) \,, \tag{2.2}$$

and, the Filling Factor (FF) is defined as follows:

$$\text{FF} = \frac{\text{total volume of all the ionized regions}}{\text{simulation volume}} \tag{2.3}$$

In the previous work in [81] and in the initial portion of this paper, we consider the same simulations as the fiducial reionization model presented in [41]. In [41], the $N$-body simulations were performed with the $CUBEP^3M$ code [94], which is based on the PMFAST algorithm, and were run as a part of the PRACE4LOFAR project (PRACE projects 2012061089 and 2014102339) [95]. The simulation cubes in these runs were of $500\,h^{-1}$Mpc = 714 cMpc in length along each side. They had $6912^3$ particles of mass $4.0 \times 10^7 M_\odot$ on a $13824^3$ mesh. This was then down-sampled to a $600^3$ grid for reionization modeling. For each redshift output of the $N$-body, halos were identified using a spherical overdensity scheme. Minimum halo mass ($M_{h_{min}}$) that was considered here is $2.02 \times 10^9 M_\odot$.

In the later part of this paper, we simulate new H I 21-cm cubes for our realization of the EoR using $1024^3$ particles of mass $1.089 \times 10^8 M_\odot$ on a $2048^3$ mesh grid. The final reionization maps were generated by coarsening the initial mesh grid with a factor of 8 resulting in a $256^3$ grid volume. The smallest halo that was considered in this simulation has a mass of $1.089 \times 10^9 M_\odot$. The simulation cube has a length of 143.36 cMpc along each side. In this realization of the simulations, we kept the mean free path constant at $R_{mfp} = 20.0$ cMpc and tuned the $N_{ion}$ values for every redshift to keep the reionization history (represented by the neutral fraction $x_{HI}$) consistent with the Fiducial model used in [41], as mentioned earlier.

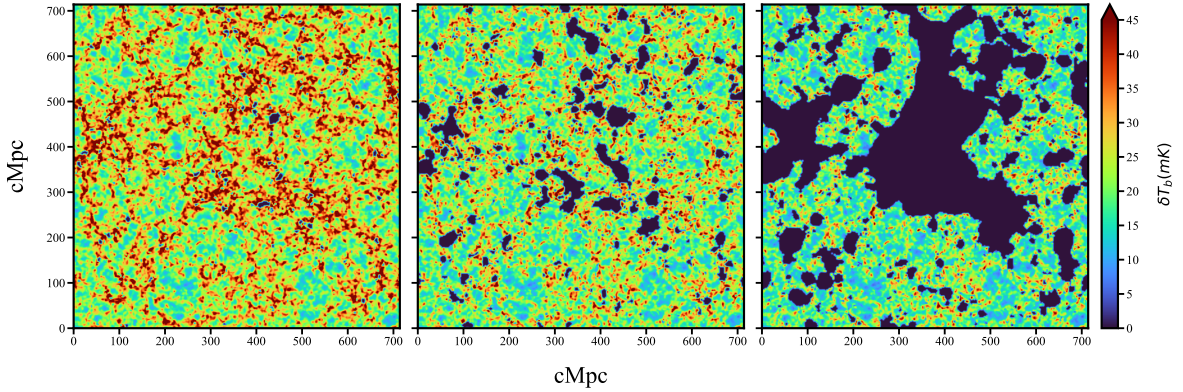

**Figure 1**. Slices from the simulated and down-sampled H I 21-cm image cube of $300^3$ grid units generated at neutral fractions *[from left to right]*: $\bar{x}_{HI} \simeq 0.9$, $\bar{x}_{HI} \simeq 0.75$ and $\bar{x}_{HI} \simeq 0.5$. The development of reionization can be clearly observed as the number of "zero-intensity" pixels, representing the ionized regions starts to grow as we go later stages of reionization i.e. at lower neutral fractions.

### 2.3 Adding observational effects to the simulated 21-cm maps

#### 2.3.1 Impact of low-resolution

The future SKA1-Low telescope will typically have a lower resolution than the simulations presented in [41], which has a resolution of ∼ 1.19 cMpc along each side of the simulation volume. To investigate the impact of this lower resolution of observations on the derived LCS, we coarse-grid the simulated signal maps and down-sample them to a resolution of 2.38 cMpc along each side of the box.

This down-sampling is done in the following manner, we first randomly choose one of the 8 neighbouring cells in the simulated signal maps and only this one out of 8 cells is kept in the down-sampled data. Through this method, our original simulated maps of $600^3$ grids turn into $300^3$ grids maps, while keeping the physical volume of the maps the same (i.e 714 cMpc) for both cases. A pictorial representation of one of the slices from this lower-resolution map can be seen in Figure 1.

Radio interferometric observations do not have zero-length baselines or visibilities with $k = 0$ wave number. This results in the fact that the images obtained from radio interferometric observations will not have absolute flux calibration. This in turn leads to an image that has its mean subtracted. Hence, the ionized pixels in our simulations (i.e $\delta T_b = 0$) will be mapped to the minima of a mean subtracted field (i.e $\delta T_b = -\bar{\rho}_H$). To mimic this behaviour we subtract the mean from all of the simulated maps considered here. Figure 2 shows that after down-sampling and mean subtraction the reionization history for our maps remains the same.

#### 2.3.2 Impact of Noise and Beam

A thorough study on SKA1-Low-like noise done by [96] and [97] showed that the noise can be well estimated using a Gaussian random field. The RMS of such a field is expressed

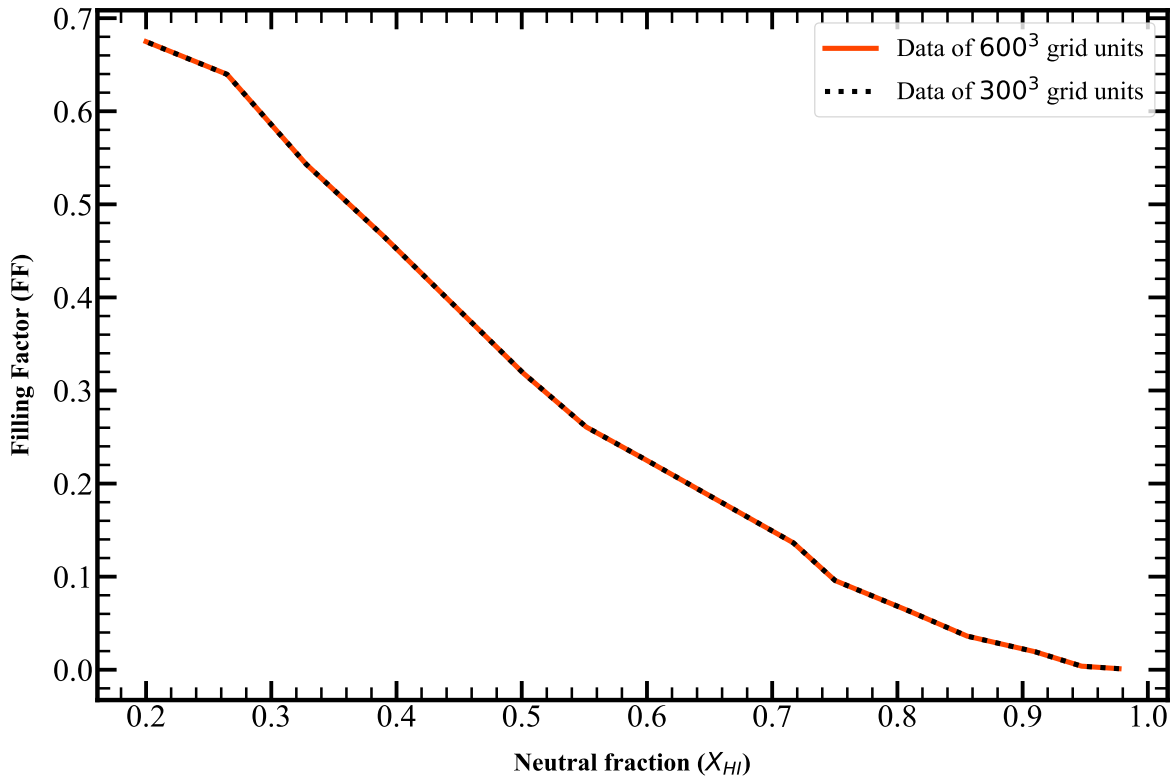

**Figure 2**. FF vs $\bar{x}_{HI}$ plots are shown for the data cubes of $600^3$ grid units and $300^3$ grid units. It can be seen that the time evolution for Filling Factor (FF) remains unchanged even after the resolution of the maps are lowered and the mean is subtracted.

through the radiometer equation [97]:

$$\sigma = \frac{\sqrt{2} K_B T_{sys}}{\epsilon A_D \sqrt{\Delta \nu t_{int}}} \ . \tag{2.4}$$

The $T_{sys}$ is the telescope system temperature, $K_B$ is the Boltzmann constant, $\epsilon$ is the antenna efficiency and it is dependent on the frequency of observation. $A_D$ is the physical area of each dish, $\Delta \nu$ and $t_{int}$ are the frequency resolution and the telescope integration time respectively.

The authors of [96] have estimated the distribution of ionized bubbles in the 21-cm maps, and have considered a noise of rms 2.82 mK for their analysis. As our work focuses on the largest ionized region at any stage thus we start with a higher noise rms i.e 3.10 mK. We simulate Gaussian random noise with the rms values ranging from 3.10 mK to 9.30 mK and study its impact on the recovered LCS behaviour.

Further, the image generated by any radio interferometer will have the sky brightness convolved with the beam of the telescope. This convolution process has a smoothing effect on the image and it further degrades the image resolution. To imitate this effect we use a 3D Gaussian kernel with varying smoothing length-scales starting from 7.14 cMpc to 14.28 cMpc along each side of the simulation volume. In Figure 3, a visual representation of these effects is shown in the three panels.

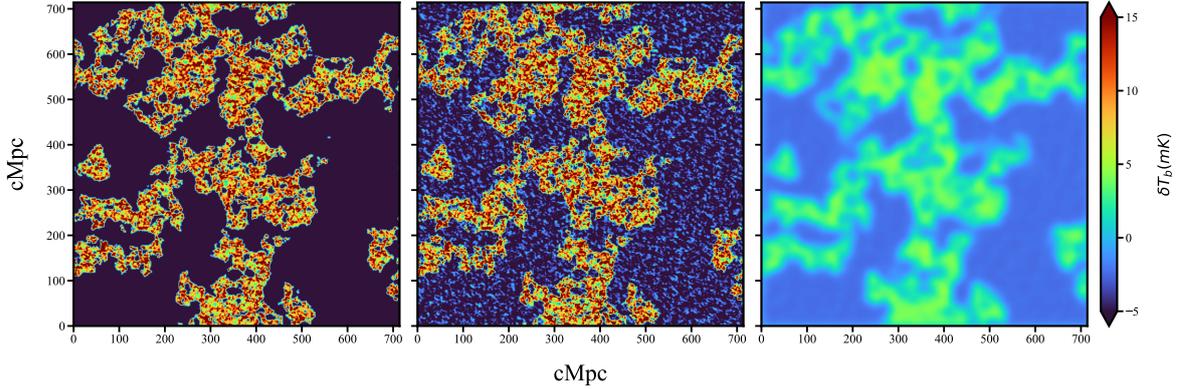

**Figure 3**. Pictorial representation of the different slices from an H I 21-cm image cube of $\bar{x}_{HI} = 0.2$. *Left:* The simulated image cube after down-sampling and mean subtraction. It can be seen from the colorbar on the right that the zero level has been shifted to a lower negative value. *Centre:* Representation of the same image slice after the addition of a Gaussian random noise of $N_{rms} = 3.10$ mK. Random fluctuations in the 21-cm map is observed. *Right:* The observed 21-cm signal cube slice after smoothing with a 3D Gaussian kernel on top of the noisy image map. It can be observed that smoothing introduces some de-noising effect.

### 2.4 Simulating realistic SKA 21-cm maps

To analyse the robustness of LCS, we further simulate a synthetic radio observation as per the characteristics of the upcoming SKA1-Low. The schematic diagram for the 21cmE2E-pipeline is shown in Figure 4. These simulations are created using OSKAR[7] software for SKA1-Low. Common Astronomy Software Application (CASA[8]) is used to create 21-cm images from the visibility data generated from OSKAR. In this work, we have considered an observation for half an hour (±0.5 HA), with the phase centre at $\alpha$ =15h00m00s and $\delta = -30°$. The integration time for this observation is 120 seconds. One of the axes of the simulated coeval 21-cm cubes was simply labeled to be as the Line-of-sight axis (LoS) or frequency axis and the cubes were divided into slices according to their spatial resolution. Each of these slices has a frequency label according to its comoving distance from the observer. These slices of the simulated 21-cm maps are then converted from comoving coordinates to angular coordinates in the sky plane before feeding into OSKAR. The 21cmE2E-pipeline outputs are then stacked slice by slice according to their frequency label to generate the final image cube.

#### 2.4.1 Telescope Model

The Square Kilometer Array (SKA-1 Low) is a powerful radio telescope that is expected to have the necessary sensitivity (Signal-to-Noise Ratio (SNR)) to create tomographic images of the 21-cm signal from the EoR [33]. The SKA-1 Low will operate in the frequency range 50 to 350 MHz and will be located at the Murchison Radio-astronomy Observatory (MRO) in Western Australia. The array is designed to have a dense core, spiral arms, and 512 stations with a maximum baseline length of 65 km. Each station will have 256 antennas and

---

[7]https://github.com/OxfordSKA/OSKAR/releases
[8]https://casaguides.nrao.edu/index.php?title=Main_Page

– 9 –

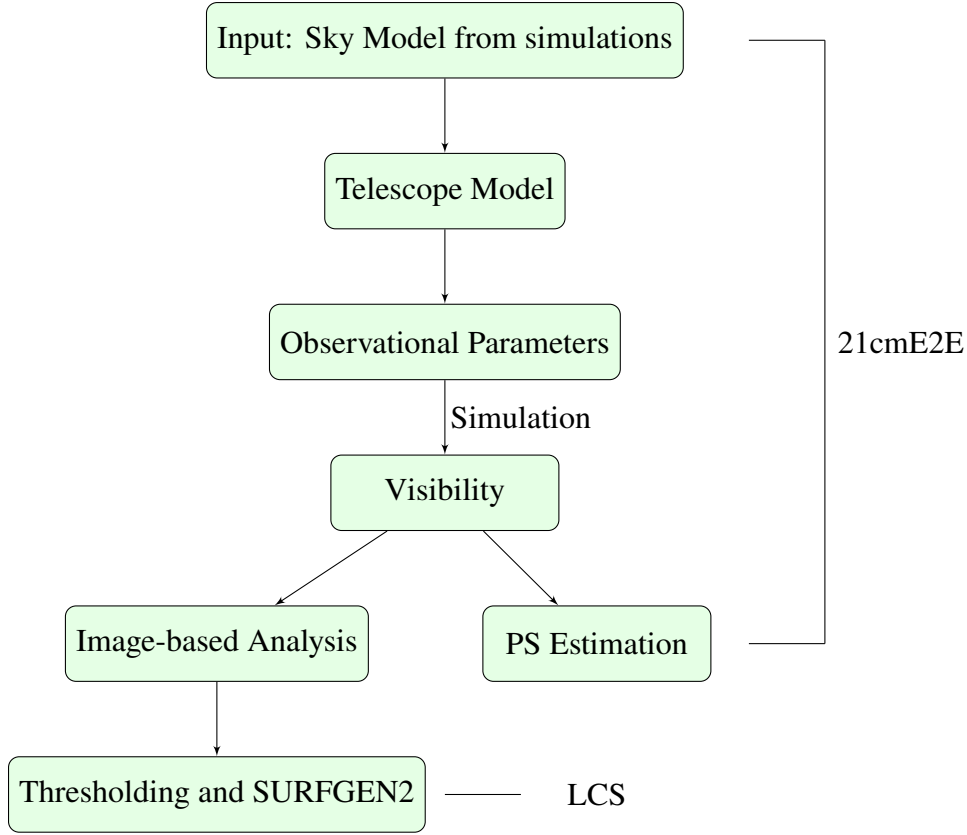

**Figure 4**. Schematic diagram of the 21cmE2E-pipeline [82] used for analysing the H<small>I</small> 21-cm maps from the future SKA observations.

a diameter of less than 40 m. The compact core of the array, which is composed of ∼ 85% of all stations within a baseline of 2 km, is designed to maximize the sensitivity of the target large-scale signal. This is illustrated in Figure 5.

### 2.4.2 OSKAR

The OSKAR package contains a number of applications for simulating radio interferometric observations. In order to produce simulated observations from aperture array-using telescopes, such as those foreseen for the SKA, OSKAR was principally developed. OSKAR reproduces digital beamforming for an aperture array of a SKA-sized station. The simulator was developed as a versatile study platform to examine alternative computational beamformer processing strategies and their impact on the output beam quality [98]. The radio interferometer measurement equation as mentioned in [99] is used in OSKAR to generate the visibility data from the sky model. The "end-to-end beamforming simulation" function of OSKAR generates antenna outputs for each element of the telescope array using a point-source sky model. Because this mode simulates a full-size, authentic SKA station, it is possible to replicate beam tracking and add several beams per level, noise, and other time-variable effects.

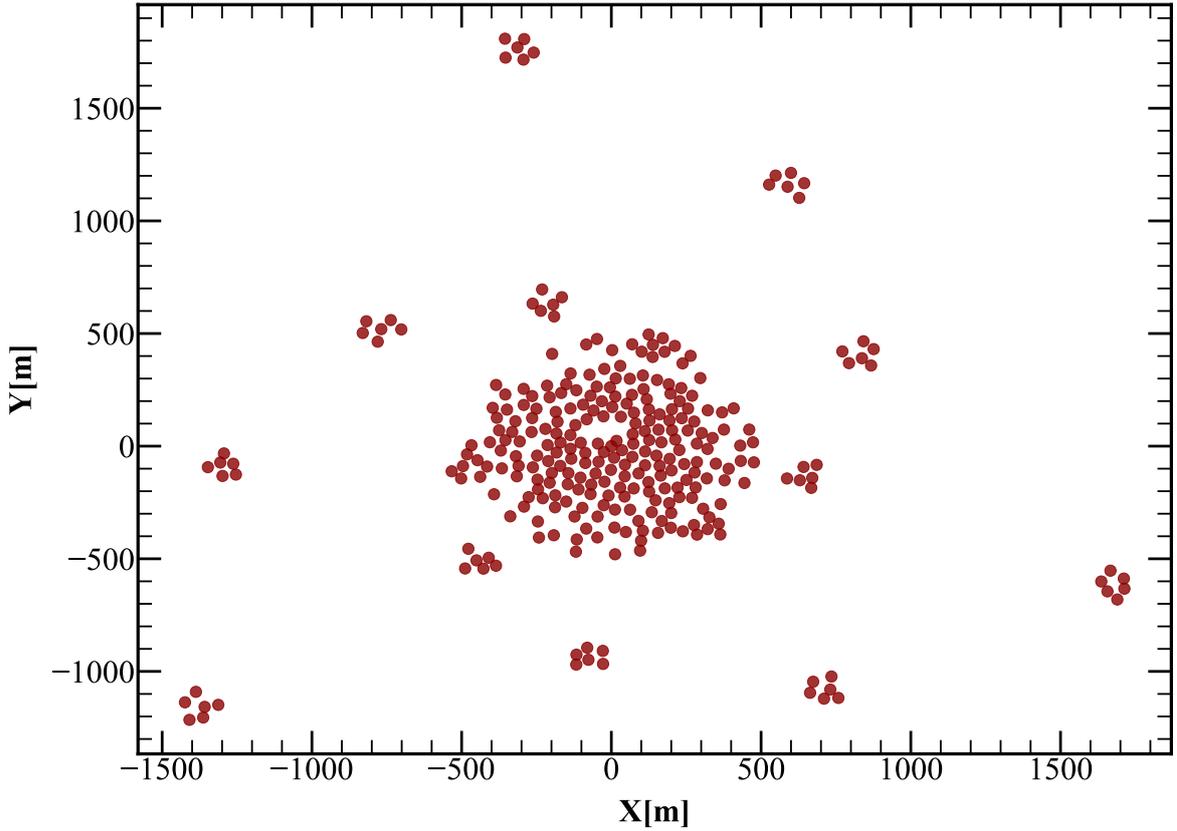

**Figure 5**. A simple layout of the telescope model used in this work. The stations residing at 2km from the central station are operational.

Using the Rayleigh-Jeans equation as mentioned in equation 2.5, we convert the semi-numerically simulated Hı 21-cm maps to maps of specific intensity. This enables us to use these maps as an input to the SKA pipeline, namely, the 21cmE2E-pipeline.

$$S = \frac{2k_B T}{\lambda^2} \Omega, \quad (2.5)$$

here $S$ denotes the specific intensity, $k_B$ is the Boltzmann constant and $\lambda$ is the wavelength of the corresponding central frequency of the observation. The calculation of the frequency resolution requires a conversion from comoving coordinates to angular coordinates. This was done using the formalism as mentioned by the authors of [96] and are described in equation 2.6 and equation 2.7.

$$\Delta\theta = \frac{\Delta x}{D_c(z)}, \quad (2.6)$$

where $\Delta x$ is the spatial resolution of the image slice in comoving units, $D_c(z)$ is the comoving distance to redshift $z$, and

$$\Delta\nu = \frac{\nu_0 H(z) \Delta x}{c(1+z)^2}, \quad (2.7)$$

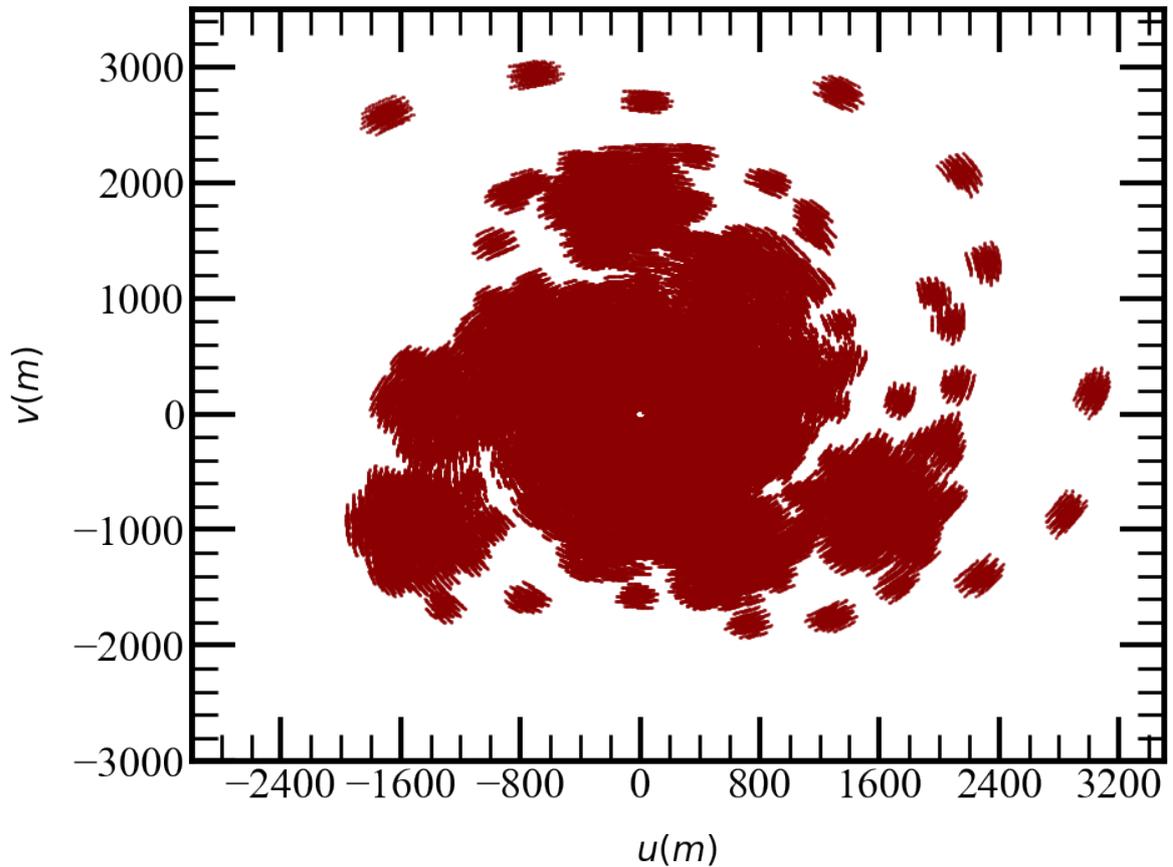

**Figure 6**. The *uv*-coverage for the simulated SKA1-Low telescope using OSKAR for an observation time of 30 mins.

where $c$ is the speed of light, $H(z)$ is the Hubble parameter at redshift $z$, and $\nu_0$ is the rest frequency of the 21-cm line. The simulated signal cube size is the same for all datasets, but the angular resolution was changed due to the different redshift (i.e. observing frequency). For instance, a slice from the simulated signal cube of 714 cMpc$^3$ at redshift $z = 7.221$ has an angular resolution of 0.015 deg (54 arcsec ), and frequency resolution of 125 KHz. The transformed images were used as the sky model for the OSKAR simulator and then the visibility data corresponding to these images were obtained which incorporates the telescope synthesized beam-effects using the van Cittert-Zernike equation as shown in equation 2.8 [52]. After the synthetic observation, the angular resolution was changed. The angular resolution depends on the maximum baseline of the interferometer and the central frequency of the observation. The angular resolution of the final images was changed due to the different central frequencies. For example, at redshift $z = 7.221$ the resulting map has an angular resolution of 0.00833 degree (or 30 arcsec). The *uv*-coverage of the simulated telescope for 30 minutes of observation is shown in Figure 6.

$$A(l, m, \nu)I(l, m) = \int \int_{sky} V(u, v, \nu) exp[2\pi i(ul + vm)] du dv, \qquad (2.8)$$

here $A$ denotes the primary beam pattern, $I$ is the sky intensity and $V$ is the observed visibility from the interferometer. As this work does not deal with the primary beam correction, we consider $A(l, m) = 1$. OSKAR employs a Fast Fourier Transform scheme to calculate the visibility using FFTW [100]. Hence, it is reasonable to use a simulation volume of the order of $2^{3n}$ in grid units to avoid the effect of zero padding. The effects of this zero padding may result in the creation of artefacts on the maps that is discussed in Section 4.

In this study, we only consider the compact core of the array, within a baseline of 2 km from the central station to create the visibility data-set. The shorter baseline is more sensitive to the target large-scale signal, that is why we are restricting our simulation to the 2 km core baseline.

In this particular work, we only consider the effect of Point Spread Function (PSF) on the Largest Cluster Statistics (LCS) of H I 21-cm maps and the inference drawn therein. However, the effect of thermal noise is not taken into account in this implementation of OSKAR simulation.

### 2.4.3 Synthesis imaging using SKA 21cmE2E-pipeline

One of the major aims of this work is to study the impact of the array synthesized beam on the recovered LCS information from the redshifted 21-cm maps from a simulated SKA1-Low observation. One of the widely used deconvolution algorithms for imaging astronomical sources is CLEAN, which has an implementation with the CASA software as well. The CLEAN algorithm allowed for the synthesis imaging of complex structures even when the Fourier plane had relatively weak coverage, as is the case with partial earth rotation synthesis or arrays of few antennas. The sky image is considered to be a collection of point sources in CLEAN. The algorithm works by iteratively calculating the brightness and location of every point source in the sky from the image. The measured visibilities are weighted and plotted on uniformly spaced grid points to determine the Fourier transform of the measured visibilities from OSKAR. The terms uniform and natural weighting in CLEAN, relate to two extreme weighting regimes. Natural weighting equalises the weights of all measured values and adds them together. Natural weighting works by emphasising on the area of the visibility plane with the most measurements. It is to be noted that the Fourier transform of the sampling function multiplied by the weight kernel should be treated as the effective dirty beam when implementing a weighting scheme. Contrarily, in uniform weighting, the visibilities are weighed according to the spatial density of the observed visibilities prior to screening. Although the PSF is smaller, as a result of this weighting, the sidelobe power is larger. The robust weighting technique [101] is an attempt to integrate natural and uniform weighting. To support high-bandwidth, high-dynamic-range imaging, an iterative weighting method that minimises PSF variation across frequencies while maximising sensitivity is discussed in [102]. In [52], the specific gridding procedure, different weights, and their effects are covered in detail.

Deconvolution is the process by which we construct the true sky brightness distribution from the observed data. The main idea behind deconvolution is the subtraction of the instrument's PSF from the given dirty image. The CLEAN algorithm is commonly used for this purpose and consists of two parts: the major cycle, which converts the visibility data into image data, and the minor cycle, which performs the deconvolution steps that separate the sky brightness distribution from the instrument's PSF. There are many algorithms for the deconvolution steps, such as *Hogbom, Multiscale, Clark, and Clarkstokes*. We first applied the Hogbom method to reconstruct the true sky emission which is represented in the top right panel of Figure 7. However, this method is not extensive in reconstructing the images from the original ones. To mitigate this problem, we applied the multiscale algorithm (see more [103]), which assumes that the sky emission is a combination of Gaussian functions rather than delta functions. This method is useful for images showing extended emission from the true sky. A pictorial representation of the Multiscale cleaned image with natural weighting is shown in the bottom left panel of Figure 7. In the context of radio interferometry, the CLEAN algorithm is considered greedy because it aims to identify and remove the brightest point sources from the image as quickly as possible, without paying much attention to the distribution of the sources in the image. The algorithm starts by identifying the brightest point in the dirty image and then subtracting a scaled version of the PSF at that location. This process is then repeated for the next brightest point, and so on, until a stopping criterion is met. This can lead to over-subtraction of the sources and loss of information, especially in the case of extended sources. Our main goal is to recover the largest ionized region for further statistical analysis. To improve the CLEAN process further, the Multiscale algorithm with Briggs weighting (with robust parameter set to a value close to natural weighting) was used to recover the images as shown in the bottom right panel of Figure 7. This method has some advantages, such as the ability to reconstruct larger bubble sizes of an ionized region, resulting in higher sensitivity for extended sources. However, the main disadvantage is that the resolution of the images will be poorer. After deconvolution, the output sky model is restored by a Gaussian function representing the instrumental resolution specified by the PSF main lobe but without the side lobes. This deconvolution method creates another issue, the bimodal nature of the 21-cm image histogram is lost. Therefore, the threshold finding for the LCS analysis becomes challenging.

## 3 Methodology

### 3.1 Percolcation Transition and LCS

At the beginning of reionization ($z \sim 15$ [104]), the newly formed luminous objects start to emit copious amount of ionizing photons and gradually ionize the surrounding IGM which was mostly abundant with neutral hydrogen. With reionization progressing, small pockets of ionized regions start to appear and they grow in size and numbers. At some point in time, these regions start to overlap and eventually merge to form a singly connected ionized region. This "phase transition" when numerous small ionized regions abruptly merge together to form a large connected ionized region that extends through out the IGM is called "*percolation transition*" [105, 106]. In the context of this work, we identify the percolation transition when the largest ionized region stretches from one end of the simulation volume to the other end and it becomes formally infinitely extended due to periodic boundary conditions.

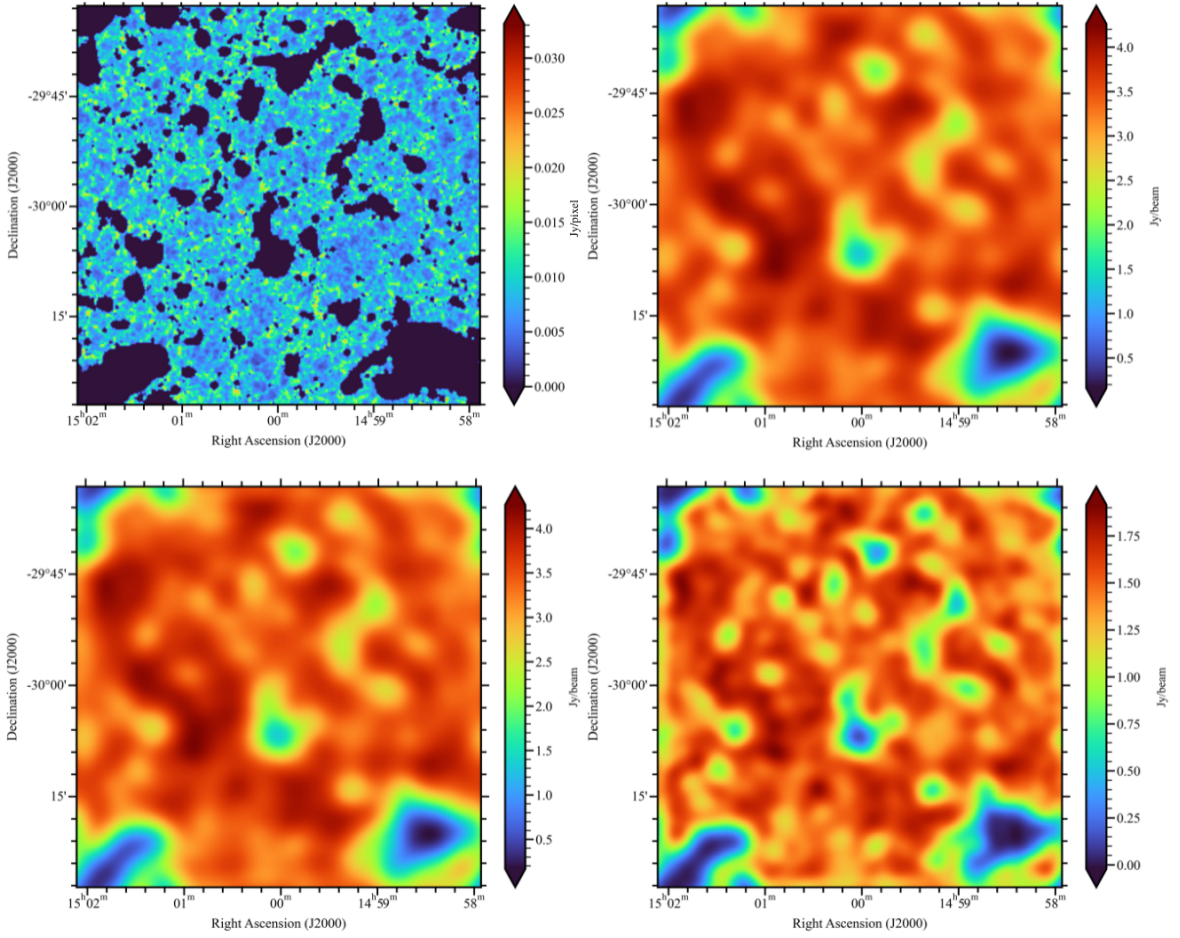

**Figure 7**. Effects of different deconvolution algorithms employed via the 21cmE2E-pipeline on an H I 21-cm image cube slice of $\bar{x}_{\rm HI} = 0.55$ are shown. *Top Left:* The Simulated 21-cm image slice of 256 grid units at $\bar{x}_{\rm HI} = 0.55$ used as the input true-sky model in the 21cmE2E-pipeline. *Top Right:* Slice of a H I 21-cm image cube at $\bar{x}_{\rm HI} = 0.55$ obtained after performing Hogbom CLEAN algorithm. *Bottom Left:* Slice of the same simulated observational H I 21-cm signal map cleaned using Multiscale deconvolution with natural weighting. *Bottom Right:* Slice of the mock observational H I 21-cm image slice at the same neutral fraction obtained after deconvolving with Briggs weighting scheme with Robust = 0.5. It can be seen that as we go from Hogbom to Multisclae deconvolution and subsequently set the robust parameter as 0.5, we move closer towards the true-sky model image.

In this particular work, which is a follow-up of [70, 81], we follow the largest ionized region (LIR) with the progress of reionization. For our study we use a novel statistic called the "Largest Cluster Statistic" (LCS) [105–107] to track the percolation process and it is defined as follows:

$$\text{LCS} = \frac{\text{volume of the largest ionized region}}{\text{total volume of all the ionized regions}}. \quad (3.1)$$

From the above definition, it can be stated that LCS essentially denotes the ratio of the largest ionized volume to the total ionized volume. Hence, at the onset of the percolation transition an abrupt increase in LCS is expected. The point at which this abrupt transition takes place is defined as the percolation transition threshold. We plot LCS as a function of the mass averaged neutral fraction ( $\bar{x}_{\rm HI}$ ) to describe the development of the Largest Ionized Region (LIR) with changing neutral fraction. As the history of reionization is model dependent, we do not plot LCS against redshift; rather we plot it against $\bar{x}_{\rm HI}$ as it is a better metric to quantify the ionization state of the universe at a given redshift. The critical $\bar{x}_{\rm HI}$, on the other hand, denotes the value of $\bar{x}_{\rm HI}$ at which percolation transition takes place as small ionized regions merge to form a large ionized region that extends throughout the IGM. This leads to the volume of the LIR to grow abruptly and a sharp rise in LCS is expected. We identify the percolation transition threshold to be that point where the change in LCS is maximum. Hence, both the profile of the LCS and its critical value at the percolation transition are important metrics to evaluate the morphological development of reionization and its history. Results from [70] and [81] corroborate this finding. The distinction between outside-in and inside-out models using LCS is described in detail in our previous work [81]. In the same work, the source models which can be distinguished using LCS are also discussed in detail.

### 3.1.1 Binarization of the image cubes

To determine LCS on the brightness temperature maps, we employ a sophisticated code called SURFGEN2 [70, 71] which is an advanced version of the original SURFGEN code developed by the authors of [108–110] to study the large scale structures of the universe. SURFGEN2 not only determines the LCS, but also helps to find the topological and morphological features of the individual ionized regions using Shapefinders. The working principle of SURFGEN2 can be found in [70, 71, 81, 111].

The working of SURFGEN2 necessitates a threshold to distinguish between the neutral and the ionized pixels in the 21-cm maps. It is important to note that no partially ionized regions are considered in the SURFGEN2 code. As mentioned earlier, for an ideal HI 21-cm field, $\rho_{\rm HI}(\mathbf{x}) = 0$ can be considered to be the boundary between the ionized and the neutral regions. But, due to computational constraints we choose $\rho_{\rm HI}(\mathbf{x})$ to be a very small number ($\simeq 0.1$) as a threshold for the ideal case.

As we go to radio-interferometric images, we miss out on the zero-length baselines for physical reasons. This results in the subtraction of the mean from the 21-cm images. Hence, the threshold for these cases will be the minima of the fields (i.e $\rho_{\rm HI}(\mathbf{x}) = -\bar{\rho}$). We use this threshold on each of the 21-cm field and the results for the mean subtracted maps exactly follow that of the ideal case, as expected.

As we add Gaussian random noise and run over a Gaussian smoothing kernel to create realistic images, identification of the optimal threshold becomes a non-trivial task. Addition of noise generates random fluctuation in the pixel values that changes the brightness temperature in each of the pixels with some amount defined by the rms of the noise. This, in turn shifts the minima of the fields to a random point that is unknown to the user and some other pixel gets mapped to the minima of the field. Hence, the minima of the fields could not be used as a threshold to determine the LCS in such scenarios.

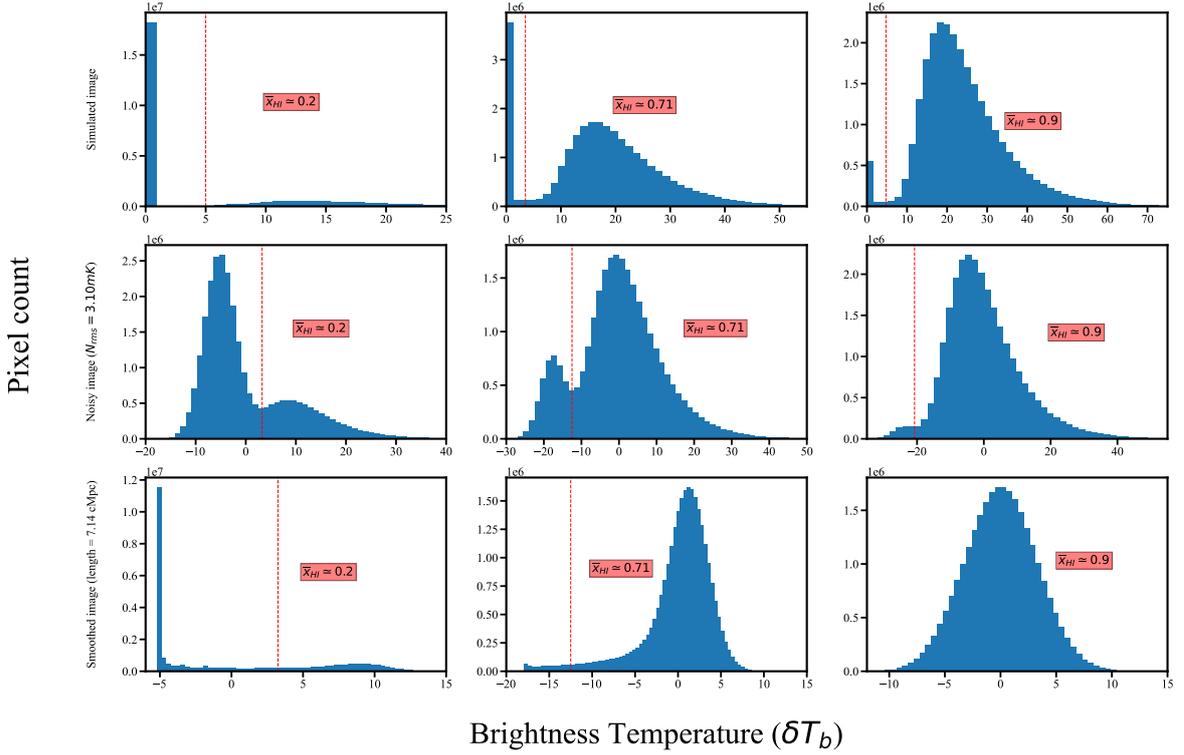

**Figure 8**. Evolution of the bimodality of the histogram of H I 21-cm field with changing neutral fraction ($\bar{x}_{\rm HI}$). [*Top panel: R-L*] Each panel represents the histogram of the 21-cm field at a fixed neutral fraction. As the universe gets ionized, the left peak, representing the ionized pixels, grows in size. On the other hand, the peak on the right, representing the neutral pixels gets smaller as the ionization fraction of the universe increases. The red vertical dashed line in each panel represents the identified threshold by the gradient descent algorithm. [*Middle panel: R-L*] Adding a Gaussian random noise introduces random fluctuation in the brightness temperature values in the 21-cm field which shifts each of the pixel intensities by a random value defined by the noise RMS. This creates a bias in the threshold identification by gradient descent. [*Bottom panel: R-L*] At a higher neutral fraction ($\bar{x}_{\rm HI} \simeq 0.9$), the gradient descent algorithm fails to identify a proper threshold as the smoothing effect of the beam washes out the bimodality of the 21-cm image histogram. The Gaussian smoothing kernel creates an averaging effect on the 21-cm maps creating regions that appear as partially ionized regions. This affects the choice of a threshold by the gradient descent, and therefore, a bias is introduced in the selection of the threshold.

The histogram of an ideal H I 21-cm field has a very sharp bimodal feature [112] that evolves with time as seen in the top panels of Figure 8. The left delta function-like peak in each panel represents the ionized pixels and the rest of the non-zero histogram represents neutral pixels. It can be seen that as reionization progresses, the delta-like peak, representing the ionized pixels grows. Alongside, the peak on the right-hand side loses its height as the neutral regions get diminished by the incoming ionizing photons. We utilize this feature and set our threshold as the local minima between these two peaks. To find the local minima, we employ the method of Gradient Descent.

To utilize the Gradient Descent algorithm, we smooth the histograms after choosing an optimum bin size for the histograms. We verify our method by running Gradient Descent on the mean-subtracted image cubes and find that the chosen threshold is in-line with the minima of the ideal fields. We apply the method of Gradient Descent to find the threshold for the Noisy and Gaussian smoothed image-cubes to calculate the LCS. We apply the same technique on the maps produced by the 21cmE2E-pipeline and calculate LCS for each of the image cubes.

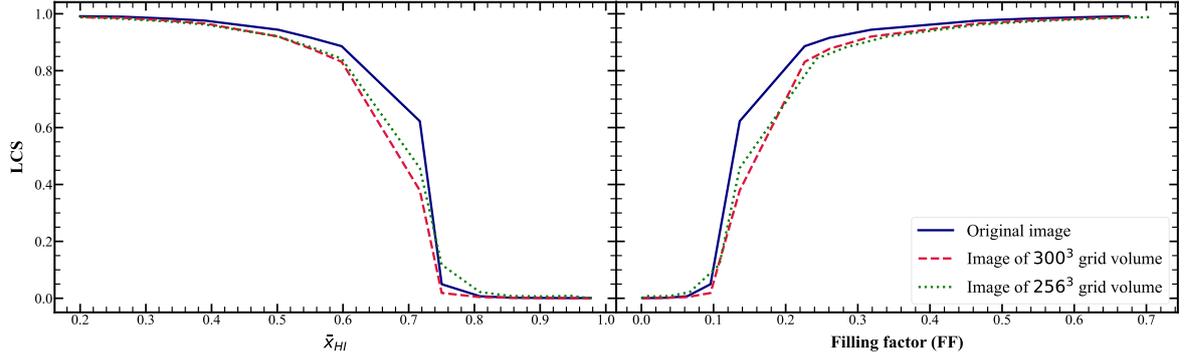

**Figure 9**. The Largest Cluster Statistic (LCS) is plotted with changing neutral fraction ($\bar{x}_{HI}$) and Filling Factor (FF) for 21-cm image cubes of different resolutions. In both panels, the solid blue line denotes the development of LCS for the original simulated data cubes of $600^3$ grid volume. The red dashed line represents the coarsened maps of $300^3$ grid volume. The green dotted line represents the data cubes of $256^3$ grid volume. It is observed that the percolation transition threshold is consistent for all three resolutions.

As we add a higher noise rms, we lose the bimodal feature of the 21-cm image histogram at higher $\bar{x}_{HI}$ and the gradient descent algorithm identifies a threshold value that is biased (shown in the middle panel of Figure 8). Similar cases happen for the Gaussian smoothing case with larger smoothing length scales as an averaging effect is employed on the maps. This effect is shown in the bottom panel of Figure 8. This necessitates the search for a better threshold-finding algorithm that is independent of the histogram of the 21-cm image cubes. We discuss these possibilities in Section 5.

## 4 Results

### 4.1 Effect of resolution on LCS

The original simulated H I 21-cm image cubes having a resolution of 1.19 cMpc$^3$ were coarsened down to a resolution half the original resolution (i.e. 2.38 cMpc along each side). It is expected that the future SKA1-Low observations will have a similar resolution [81].

We ran SURFGEN2 on the coeval image cubes of 300 grid units and estimated LCS for each of them. We observed edge/spill-over effects occurring in the synthetic maps due to the zero-padding operation in OSKAR, when the input data grid is not equal to $2^{3n}$. Alongside, other artefacts were also observed in these images. These led to a wrong/biased estimation of LCS. The CLEAN method that has been used to deconvolve the dirty beam from the image, in this case, is the Hogbom CLEAN [103]. We saw that the Hogbom CLEAN method could not properly deconvolve the beam from the image and it leaves residues of the dirty beam. Additionally, we suspect that several edge effects appear in the image when the synthesis radio imaging is done using the 21cmE2E-pipeline. This is due to the fact that as OSKAR employs an FFTW operation, it is optimal to use images having a size of $2^{3n}$. To mitigate the effect of zero padding, we rerun our simulations of 21-cm signal from EoR and generate signal maps of size $256^3$ grids (more details are discussed in Section 2). Additionally, for this set of simulated sky models, we also employ the Multiscale deconvolution CLEAN process [103] to achieve a better deconvolution of the dirty beam. It is important to note that the data cubes of $600^3$ grids and $300^3$ grids had a volume of $\sim 714$ cMpc$^3$, whereas the maps of $256^3$ grid has a volume of 143.36 cMpc$^3$.

We explore the percolation transition process by following the development of the LCS with changing neutral fraction $\bar{x}_{HI}$ in Figure 9. Alongside, we also explore the evolution of LCS with Filling Factor (FF) which has been defined in equation (2.3). Analysing LCS with $\bar{x}_{HI}$ and FF will enable us to understand the development of the Largest Ionized Region (LIR) with the changing ionization state of the universe both in mass averaged and volumetric terms.

In Figure 9 we plot the LCS at different $\bar{x}_{HI}$ and FF in the two panels. Initially, the size of the LIR is very small, hence a small value of LCS is seen. As reionization progresses, small ionized regions start to overlap and that causes the LIR to grow in size, leading to an increase in the LCS. It is observed that the LCS of the cubes of both $300^3$ and $256^3$ grid units follow the original $600^3$ grid units with changing $\bar{x}_{HI}$ and FF. As the percolation transition approaches, an abrupt change in the LCS is observed as the LIR suddenly grows in size and volume. It is also observed that the percolation transition threshold point for all three coincides at the same point. The percolation transition threshold for each of these cases is observed at $\bar{x}_{HI} \approx 0.75$ and FF $\approx 0.096$. This result is consistent with the previous findings of the authors of [70, 73, 75, 81, 113]. After percolation occurs, the individual LIRs also start to merge with one another and then form a singly connected LIR due to periodic boundary condition. At this time as the entire simulation volume consists of the LIR, we see that LCS saturates to unity.

## 4.2 Effect of Noise and Model Array Synthesized Beam on LCS

In Section 2.3.2, we discuss the prescription we follow for the addition of a Gaussian noise of varying rms values. In Figure 10, we plot the LCS with changing $\bar{x}_{HI}$ and FF for different noise rms values. For low noise level (i.e $\sim 3.10$ mK), we see that the LCS is almost unaffected and it retains the same development as seen for the case of the original $300^3$ grid volume image. It is also observed that the percolation transition threshold point is also consistent with the original image cube.

As we go to higher noise levels, we see that the LCS could no longer be computed after a certain point as the random fluctuations introduced starts dominating the image pixels. This, as a result, masks the small ionized regions that are created in the early stages of reionization ($\sim \bar{x}_{HI} \geq 0.7$). LCS could only be computed at a time when the LIR is sufficiently large enough to not get masked by the noise.

For even higher noise levels it is observed that the LCS could not be computed even for higher neutral fractions as noise starts dominating even more. Adding higher levels of

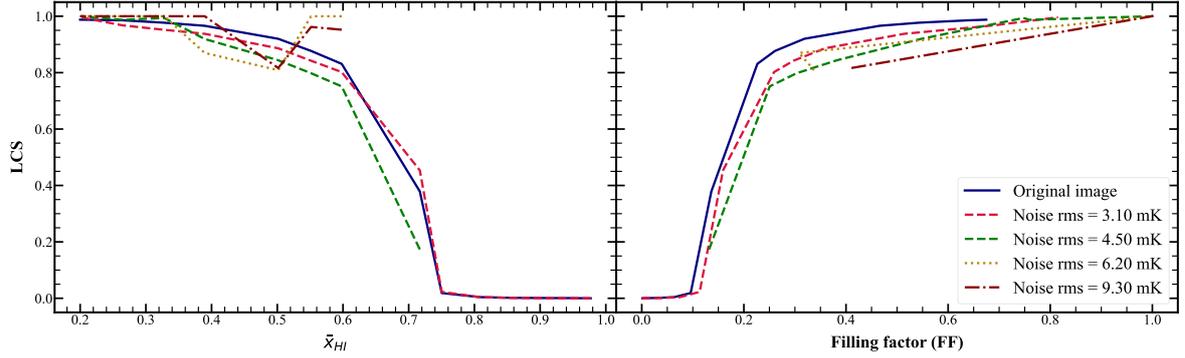

**Figure 10**. The LCS of the ionized hydrogen is displayed in the left and right panels as a function of $\bar{x}_{HI}$ and the filling factor (FF), respectively. The LCS for a noise rms of 3.10 mK follows the LCS of the original image cube. But as the noise rms goes higher, the LCS could not be computed. For extremely high noise rms cases, the gradient descent algorithm fails to binarize the 21-cm field. Hence, biased LCS results are achieved.

Gaussian noise pushes the histogram of the 21-cm maps to a more Gaussian type and the bimodal feature is lost. Hence, the method of gradient descent fails to identify the boundary between the ionized and neutral pixels. This leads to a biased computation of LCS at very high noise levels and they can be observed in the case of $N_{rms}$ = 6.20 mK and $N_{rms}$ = 9.30 mK in Figure 10.

To create mock observational effects of the array synthesized beam, we simulate a Gaussian smoothing kernel of varying FWHM as discussed in Section 2.3.2. We plot the LCS by changing $\bar{x}_{HI}$ and FF to see the impact of the beam on the detection and development of the LIR in Figure 11. We start with a Gaussian smoothing kernel having a smoothing length-scale of 7.14 cMpc and go all the way up to a smoothing length-scale of 21.42 cMpc.

For a relatively smaller smoothing length scale of 7.14 cMpc, it is observed that as the small individual ionized regions are smoothed out, numerous partially ionized regions start to appear due to the effect of averaging. These newly introduced partially ionized regions fall either into the category of ionized or neutral regions as the binarization scheme is imposed on the maps. This leads to a bias in the LCS results and that can be observed in the crimson red dotted lines in both the panels of Figure 11.

We tend to lose out on the features of LCS for a higher smoothing length scale at earlier stages of reionization. This occurs due to the fact that the averaging effect creates even more partially ionized regions in this case and the bimodality of the HI maps is lost. The method of gradient descent identifies these partially ionized regions as neutral regions and hence the volume of the LIR is zero, leading to no estimation of LCS.

For even higher smoothing length scales, we see that the behaviour of the LCS goes erratic. This is due to two major reasons.

- The partially ionized regions appearing due to smoothing shifts the histogram of the 21-cm field to a Gaussian-like nature.

- Due to the above-mentioned reason, the binarization scheme does not work properly with gradient descent algorithm and bias is introduced in the estimation of LCS.

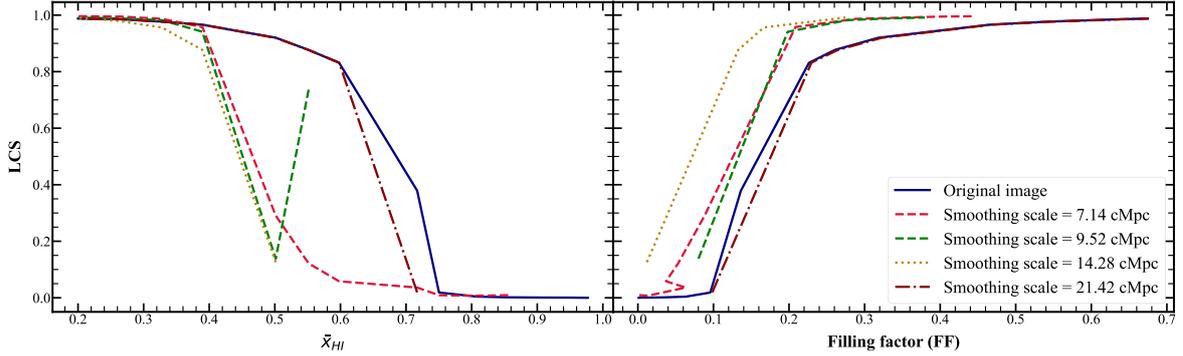

**Figure 11**. The two panels from left to right represents the variation of LCS with changing neutral fraction ($\bar{x}_{HI}$) and the filling factor (FF) respectively. Bias is introduced as the radius of the Gaussian smoothing kernel is increased. This occurs majorly due to the fact that the averaging effect caused by smoothing introduces numerous partially ionized regions which is not identifiable by SURFGEN2.

In Figure 12, we choose a constant noise rms of 3.10 mK and a constant smoothing radius of 7.14 cMpc, in-line with the future SKA observations [96] and plot LCS with varying $\bar{x}_{HI}$. We see that although the features of LCS could be computed up to a certain $\bar{x}_{HI}$, the percolation transition threshold point shifts heavily to a lower $\bar{x}_{HI}$ value. This is due to the combined effect of random fluctuation and the introduction of partially ionized regions in the H I 21-cm field. As the percolation transition threshold moves to a lower $\bar{x}_{HI}$ value, it leads to a biased interpretation of the reionization history.

### 4.3 LCS estimates through synthesis radio imaging with SKA

In section 2.4 we discuss the synthesis radio imaging procedure that we have used in this work in detail. The 21cmE2E-pipeline creates visibility data from the input sky models (which are the simulated signal maps in this case) and finally, we use CASA to make the image cubes from the visibility.

The top right panel of Figure 7 shows that the Hogbom CLEAN algorithm does not effectively deconvolve the dirty beam from the image as it assumes the sky emission to be a bunch of delta functions. This is mitigated using Multiscale CLEAN algorithm (discussed in section 2.4.3). Figure 13 compares the evolution of LCS with $\bar{x}_{HI}$ for different types of CLEAN algorithms employed on the simulated SKA1-Low observational maps. The effect of Hogbom CLEAN on LCS estimation is shown in the orange dash-dotted line in Figure 13. In the case of natural weighted Multiscale cleaned maps, LCS could not be estimated for $\bar{x}_{HI}$ values larger than 0.6. Natural weighting equalises the weights on all measured pixel values and adds them together (discussed in detail in section 2.4.3). Hence, the contrast between the neutral and ionized pixels gets reduced. This, in turn, forces a significant number of ionized regions to appear partially neutral. Effectively it diminishes the number of small ionized pixels at higher $\bar{x}_{HI}$ and subsequently, SURFGEN2 could not identify ionized clusters which leads to an estimation of zero LCS. Furthermore, as we consider a closed-packed interferometric array distribution within a 2 km radius within the central core of SKA1-Low for our mock observations, the obtained images suffer from poor resolution due to the lack of longer baselines. This, in turn, adds to the bias in the estimation of LCS at the earlier stages of reionization as the small ionized bubbles can no longer be resolved. Besides, the natural weighted deconvolution is not the optimal weighting scheme for extended sources as it leaves residues of the dirty beam on the cleaned images. The gradient descent algorithm for identification of the optimal threshold for LCS estimation is further affected as the bimodal feature of the 21-cm histograms gradually disappears (due to the presence of residues of the dirty beam) and becomes more like a Gaussian structure. This further adds to the bias in the estimation of LCS and the LCS curve thus shifts towards a lower $\bar{x}_{\rm HI}$ for percolation transition (compared to the scenario when the impact of the beam has not been considered i.e. the blue curve in Figure 13) as observed by the red dashed line in Figure 13.

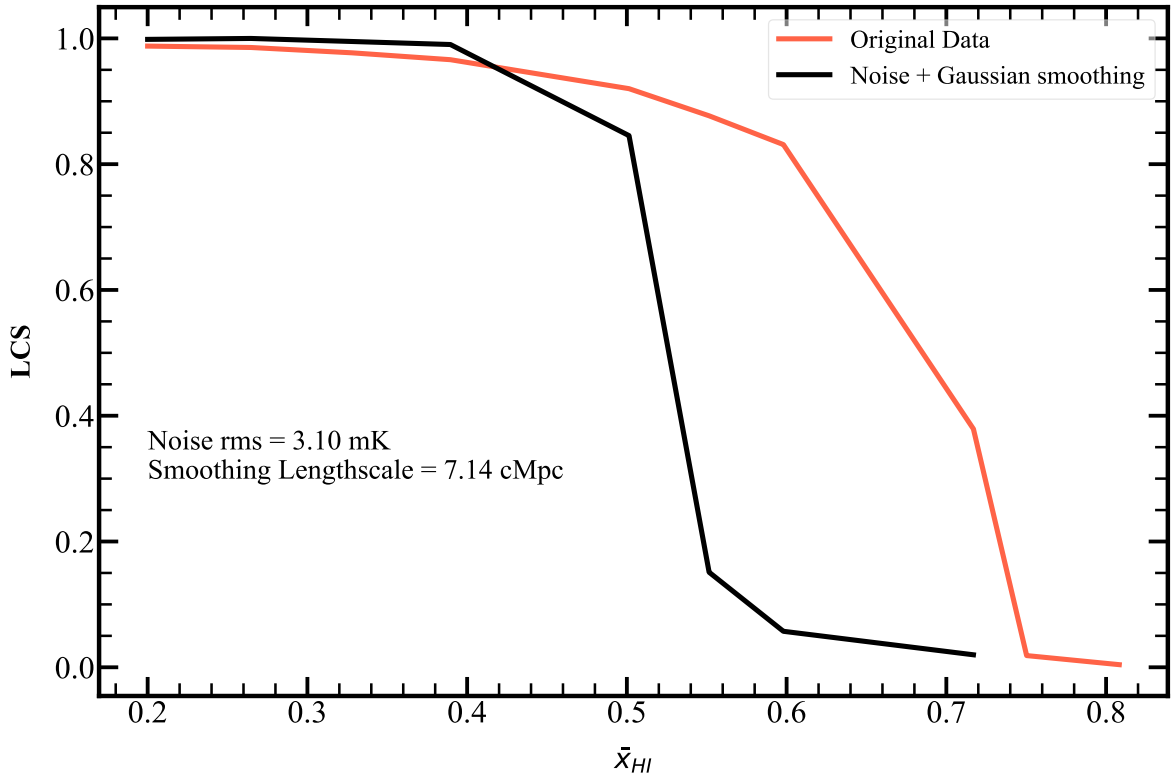

**Figure 12**. LCS is computed with varying $\bar{x}_{\rm HI}$ for a noise level of rms 3.10 mK and a smoothing length-scale of 7.14 cMpc. It is seen that the percolation process is identified at a later stage due to these effects and a biased interpretation of the reionization history is obtained.

We next use an extensive weighting scheme named Briggs weighting by setting the robust parameter in our CLEAN algorithm to be 0.5 to further push the contrast of the sky maps closer to the scenario when the impact of the beam has not been considered. This CLEAN scheme does this by enhancing the contrast between the ionized and neutral pixels. As the contrast between the neutral and ionized pixels is enhanced, the non-zero LCS estimation now could be pushed to even higher neutral fractions, $\bar{x}_{\rm HI} \approx 0.7$, than in the case of natural weighting. However, the effect of telescope resolution still persists for $\bar{x}_{\rm HI} > 0.7$ and hence LCS could not be estimated for the same reasons as mentioned for the case of natural weighted CLEAN. Even though the estimation of LCS at a lower neutral fraction $\bar{x}_{\rm HI} \leq 0.7$ could be done (as shown by the green dotted lines in Figure 13), it is still significantly biased due to the assumed Gaussian PSF and the telescope resolution. As the functional form of the PSF is unknown, the Briggs weighting scheme also leaves the residue of the dirty beam on the images and that further adds to the bias in the LCS estimation due to the sidelobes of the synthesized beam. This effectively shifts the critical $\bar{x}_{\rm HI}$ value (percolation threshold point) to later stages of reionization. Therefore, it ultimately leads to a biased interpretation of the reionization history.

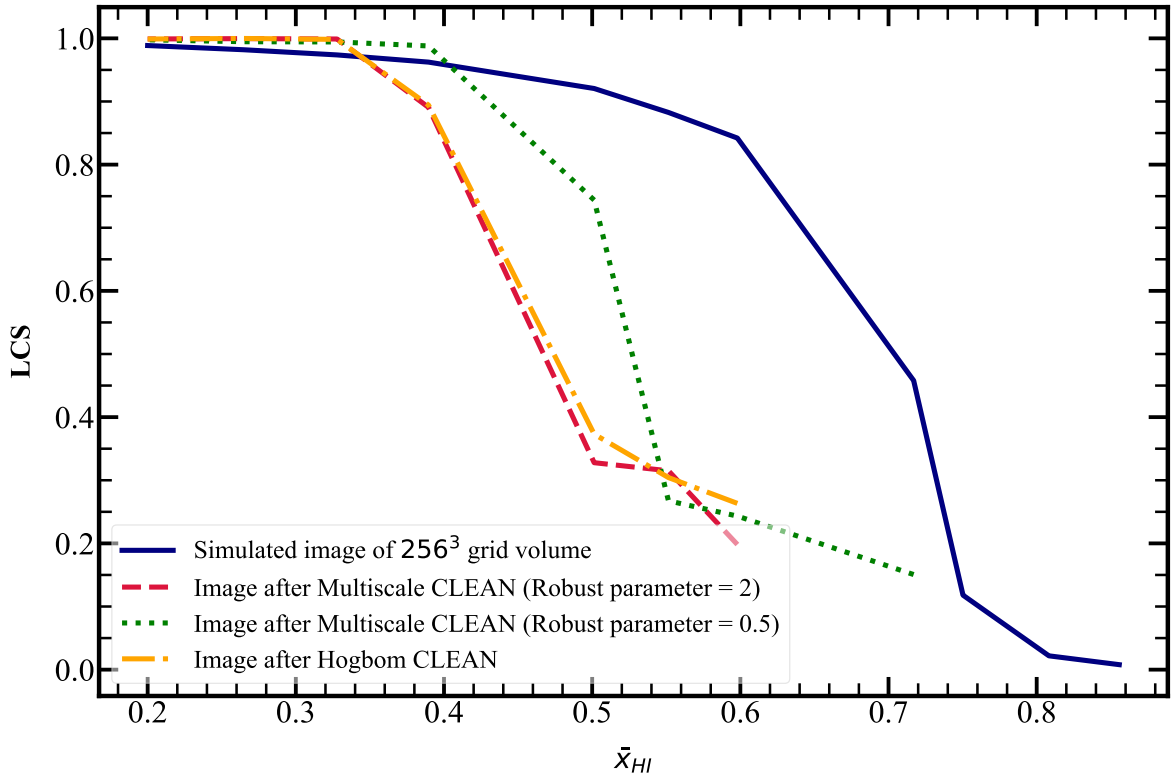

**Figure 13**. Comparison shown for the LCS obtained with changing neutral fraction after natural weighted Multiscale CLEAN and Multiscale weighted CLEAN with Briggs weighting by setting the robust parameter = 0.5. It can be seen that Briggs weighted CLEAN performs superiorly in two ways. Firstly, LCS values upto almost 0.7 could be calculated and secondly, the percolation transition is not as heavily shifted as of the natural weighted CLEAN case.

It is important to note that the results shown in Figure 13 incorporate a more accurate model of the array synthesized beam for SKA1-Low as compared to the one assumed in the case of Figure 12. We see that even without incorporating noise in the 21cmE2E-pipeline, the percolation transition threshold shifts almost as much as in the case when the effect of telescope noise and beam was implemented directly on the image plane. However, both cases will make the observer draw a biased inference that the percolation process happens at a later stage of EoR than its true value. This suggests that for the SKA1-Low to be able to create images with the required resolution to perform credible image analysis with techniques e.g.

LCS, denser *uv*-sampling is needed at longer baselines. Additionally, a longer observation period is also needed to fill the *uv*-coverage in order to reduce the impact of sidelobes and provide a more accurate estimation of the PSF.

## 5   Summary and discussion

In this work we investigate if it is possible to probe the percolation transition period of the EoR, i.e. the period when a group of small ionized regions merges together to form a large singly-connected ionized region, using a synthetic SKA1-Low observation (via 21cmE2E-pipeline) of the redshifted 21-cm signal from EoR. We use LCS to evaluate the evolution of the Largest Ionized Region (LIR) during the percolation process. To estimate the LCS, we use the SURFGEN2 algorithm that uses a Marching Cube 33 scheme to find the boundaries between the neutral and ionized regions in the H<span>I</span> 21-cm maps. The findings of this investigation can be summarized as follows:

- The determination of the threshold brightness (to differentiate between ionized and neutral regions) for a H<span>I</span> 21-cm map obtained from a realistic radio interferometric observation of the EoR is a non-trivial task. This is due to the fact that a combined effect of noise and beam on the interferometric (mean-subtracted) images shifts the brightness temperature in each pixel by a random value. The histogram of an ideal EoR 21-cm field has a very sharp bimodal feature where the two modes represent the ionized and the neutral pixels respectively. Hence, we employ a Gradient Descent scheme on the radio interferometric EoR 21-cm image histograms to binarize these fields by identifying the local minima between the two peaks of the histogram.

- We assume the presence of a Gaussian random system noise in these observations and add it to the EoR 21-cm field in the image domain with varying rms values to study its impact on the LCS analysis. For a relatively low noise rms level of 3.10 mK, we find that the LCS is unaffected and the percolation transition threshold remains unchanged. However, as we go to higher noise levels, due to higher noise fluctuations, the bimodal nature of the 21-cm histogram gets lost. This, in turn, creates a bias in the selection of the threshold by the Gradient Descent scheme. Alongside, as noise starts dominating the 21-cm image field, LCS could not be computed for higher neutral fractions ($\bar{x}_{\rm HI}$) as the contrast between the tiny ionized bubbles and their surroundings become comparable with the amplitude of the noise and it becomes difficult to identify them via SURFGEN2.

- We next mimic the impact of the array synthesized beam, by smoothing the EoR 21-cm images with a Gaussian smoothing kernel of varying length. We observe that for smaller smoothing scales, the features of the LCS could be fully recovered and the percolation threshold remains unaltered. However, as we go to higher smoothing length scales, due to the averaging effect by the smoothing kernel, a lot of patches start to appear in the smoothed image which mimics the characteristics of partially ionized regions. For images from the earlier stages of the EoR i.e. at higher neutral fractions, this effect smooths out the small ionized regions and makes them appear as neutral regions. Hence, for larger smoothing scales LCS could not be estimated at these early stages. Further, at later stages of the EoR, this averaging effect results in a bias in the determination of the threshold between the ionized and neutral regions. Therefore, one observes erratic behaviour in the LCS evolution in such scenarios.

- To investigate the impact of the interferometric array synthesized beam on the LCS analysis under a more realistic condition, we simulate observations of the EoR 21-cm signal with the upcoming SKA1-Low using the 21cmE2E-pipeline. We observe that how the dirty beam is deconvolved from the observed data has a significant impact on the resulting 21-cm images and thus on the estimated values of the LCS. We find that due to the absence of knowledge of the PSF, it is not possible to fully recover the important features of the 21-cm maps which are crucial to identify the largest ionized region, i.e. its volume and shape. We show that Hogbom CLEAN algorithm does not effectively deconvolve the dirty beam which leads to biased output as it assumes the sky emission to be a bunch of delta functions. We next demonstrate that this issue can be mitigated to some extent by using a Multiscale CLEAN algorithm. However, even a Multiscale CLEAN algorithm needs to adopt a non-trivial weighting scheme such as Briggs weighting to be able to make the SURFGEN2 algorithm work effectively on 21-cm images obtained from the early stages of the reionization. Even after adopting such an advanced CLEAN algorithm for deconvolution, still, the LCS cannot be computed for significantly early stages of the EoR i.e. $\bar{x}_{\rm HI} \geq 0.7$. This is due to the lack of the presence of long baselines in our observations, which does not allow us to resolve the smaller ionized regions in the early stages of the reionization and thus leads to a biased estimate of the critical neutral fraction when the percolation transition takes place. We conclude that we would require a much better sampling of the longer baselines in the $u - v$ plane to be able to have a less biased estimate of the LCS at the early stages of the reionization.

In this particular work, we have focused mostly on the impact of the array synthesized beam of SKA1-Low and its optimal deconvolution techniques on the LCS image analysis of the EoR 21-cm observations. Here we use the gradient descent algorithm to find the optimal threshold for LCS analysis. However, it was observed that gradient descent may not be the optimal algorithm to identify an optimum threshold. For this reason, we would like to investigate other thresholding methods in our follow-up work. Moreover, here we have not considered several other observational issues, e.g., the presence of strong foregrounds in the observed data, incompleteness of the foreground modelling, ionospheric distortions to the signal and presence of more realistic noise, all of which will make the LCS analysis even more complicated. We plan to explore these issues in our follow-up work.

# 6 Acknowledgements

SB thanks Varun Sahni, and Santanu Das for their contributions to developing SURFGEN2 in its initial phase. SKP would like to thank DST for INSPIRE fellowship. SM and AD acknowledge financial support through the project titled "Observing the Cosmic Dawn in

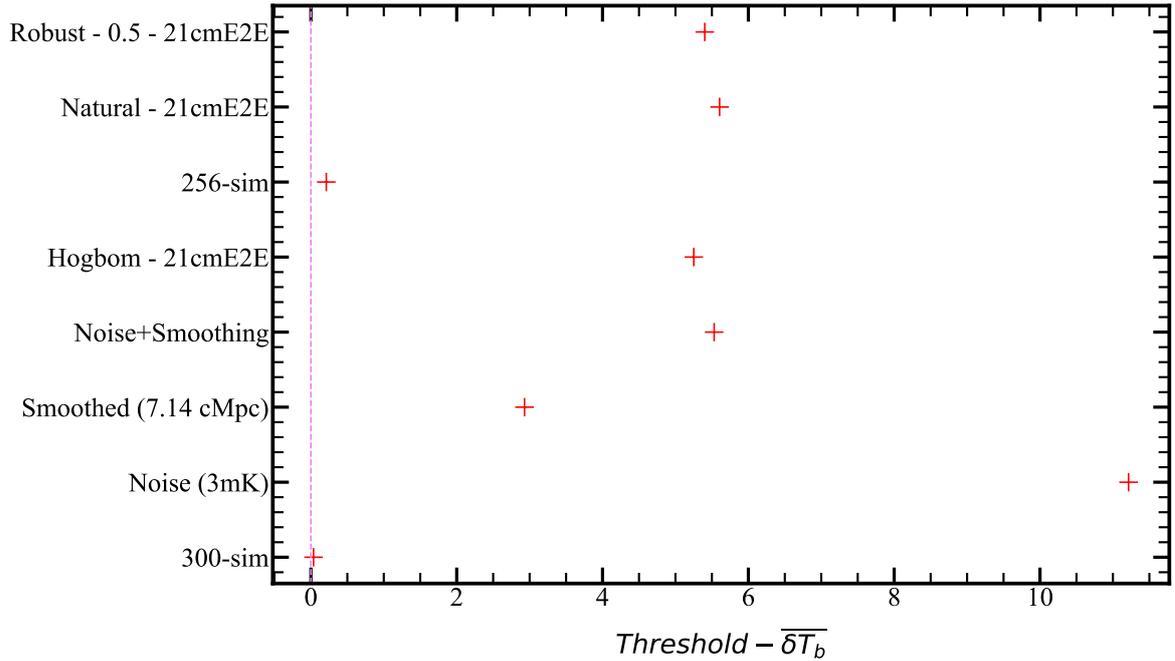

**Figure 14**. *Bottom-Up:* Threshold identified by gradient descent algorithm for different cases starting from the mean subtracted 21-cm field at the bottom-most position are shown. Subsequently, the effect of noise and smoothing directly on the image plane can also be seen. Additionally, the effect of the array synthesized beam was simulated using the 21cmE2E-pipeline and the corresponding change in the histogram is reflected by the choice of threshold for these cases. These thresholds are shown for a 21-cm signal map obtained from the late stages of reionization i.e. $\bar{x}_{\rm HI} = 0.2$.

Multicolour using Next Generation Telescopes" funded by the Science and Engineering Research Board (SERB), Department of Science and Technology, Government of India through the Core Research Grant No. CRG/2021/004025. RM is supported by the Israel Academy of Sciences and Humanities & Council for Higher Education Excellence Fellowship Program for International Postdoctoral Researchers. MK is supported by the foundation Carl Tryggers stiftelse för vetenskaplig forskning, under grant agreement 21:1376 awarded to docent Martin Sahlén. The entire analysis of the simulated 21-cm maps presented here were done using the computing facilities available with the Cosmology with Statistical Inference (CSI) research group at IIT Indore.

## A  Variation in the threshold to discern the ionized regions

A crucial aspect of the EoR 21-cm image analysis lies within the threshold selection to identify neutral and ionized pixels. In our work, we use the gradient descent algorithm to choose an appropriate threshold for our image analysis. Figure 14 shows the estimated threshold by this algorithm for different aspects of our analyses done in this paper. We plot all of the thresholds for a 21-cm map corresponding to a late stage of reionization, specifically when mass average neutral fraction $\bar{x}_{\rm HI} = 0.2$. We plot the difference between the chosen threshold identified by gradient descent and the mean of the simulated brightness temperature map for each of these cases. In the case of a simulated 21-cm map of $300^3$ grids, as shown in Figure 14, the gradient descent identifies the threshold very accurately; ergo, the difference is almost zero. After the map was corrupted using a Gaussian noise, the histogram of the 21-cm image changed significantly (as shown in the centre left panel of Figure 8) as random fluctuations arose in the brightness temperature value of the image pixels. Due to this change, the identified threshold differs from that of the simulated 21-cm map. A similar case is observed for the case of the 21-cm map smoothed with a Gaussian kernel of FWHM of 7.14 cMpc as shown in the bottom left panel of Figure 8. Due to the averaging effect introduced by the Gaussian smoothing kernel, many pixels in the 21-cm map appear as partially ionized pixels, resulting in a distortion in the histogram. This, in turn, creates difficulty in identifying the threshold using a gradient descent algorithm as we go to higher neutral fractions i.e. earlier stages of reionization. Due to the aforementioned reason, a combined impact of the noise and smoothing affects the histogram of the 21-cm image and the chosen threshold thereafter.

The edge effects and other artefacts introduced on the image created via the 21cmE2E-pipeline using Hogbom CLEAN severely corrupt the histogram of the 21-cm map. Naturally, this, in turn, affects the choice of threshold using gradient descent. The bias introduced by the pipeline imposes a bias on the threshold as well, shown in Figure 14. To mitigate the effects of zero padding and other artefacts, H I 21-cm image cubes of $256^3$ grids were simulated. The threshold chosen for one such cube of $\bar{x}_{\rm HI} = 0.2$ is shown in Figure 14. These maps were then cleaned using Multiscale deconvolution CLEAN with robust parameters set to 2 and 0.5 to deconvolve the dirty beam from the maps. However, even though these CLEAN algorithms are advanced, residues of the dirty beam remain on the cleaned maps and impose a bias on these images. Subsequently, when the gradient descent algorithm is applied to these images, the chosen threshold contains this bias. This, in turn, leads to a biased estimation of LCS and the inferred reionization history obtained, therefore, is heavily biased.